\documentclass[twocolumn,secnumarabic,,amsmath,amssymb, nobibnotes, aps, prd, superscriptaddress]{revtex4-1}

\usepackage{amsmath}
\usepackage{graphicx}
\usepackage{dcolumn}
\usepackage{floatrow}
\usepackage{xcolor}

\begin{document}
\title{Plasmonic-based gas sensing with graphene nanoribbons}
\author{Kaveh Khaliji}
\email{khali161@umn.edu}
\affiliation{Department of Electrical and Computer Engineering, University of Minnesota, Minneapolis, Minnesota 55455, USA}
\author{Sudipta Romen Biswas}
\affiliation{Department of Electrical and Computer Engineering, University of Minnesota, Minneapolis, Minnesota 55455, USA}
\author{Hai Hu}
\affiliation{Division of Nanophotonics, CAS Center for Excellence in Nanoscience, National Center for Nanoscience and Technology, Beijing 100190, P. R. China}
\affiliation{University of Chinese Academy of Sciences, Beijing 100049, P. R. China.}
\author{Xiaoxia Yang}
\affiliation{Division of Nanophotonics, CAS Center for Excellence in Nanoscience, National Center for Nanoscience and Technology, Beijing 100190, P. R. China}
\affiliation{University of Chinese Academy of Sciences, Beijing 100049, P. R. China.}
\author{Qing Dai}
\affiliation{Division of Nanophotonics, CAS Center for Excellence in Nanoscience, National Center for Nanoscience and Technology, Beijing 100190, P. R. China}
\affiliation{University of Chinese Academy of Sciences, Beijing 100049, P. R. China.}
\author{Sang-Hyun Oh}
\affiliation{Department of Electrical and Computer Engineering, University of Minnesota, Minneapolis, Minnesota 55455, USA}
\author{Phaedon Avouris}
\affiliation{IBM T. J. Watson Research Center, Yorktown Heights, New York 10598, USA}
\author{Tony Low}
\email{tlow@umn.edu}
\affiliation{Department of Electrical and Computer Engineering, University of Minnesota, Minneapolis, Minnesota 55455, USA}

\begin{abstract}
The main challenge to exploiting plasmons for gas vibrational mode sensing is the extremely weak infrared absorption of gas species. In this work, we explore the possibility of trapping free gas molecules via surface adsorption, optical, or electrostatic fields to enhance gas-plasmon interactions and to increase plasmon sensing ability. We discuss the relative strengths of these trapping forces and found gas adsorption in a typical nanoribbon array plasmonic setup produces measurable dips in optical extinction of magnitude 0.1 \% for gas concentration of about parts per thousand level.
\end{abstract}

\maketitle

\section{Introduction}
Two-dimensional (2D) materials have garnered considerable attention as platform for gas sensing, owing to their large surface to volume ratio which renders their electronic properties very sensitive to environmental influences such as adsorbed molecules or external field \citep{yang2017, liu2017, wang2016}. Thus far, gas sensing based on extended 2D materials has relied on schemes involving changes in conductance \citep{schedin2007, yim2016, zhan2014, abbas2015}, surface work function \citep{nomani2012, qazi2008}, or their contact-barrier heights \citep{singh2014, quang2014, liu2014}, with the main underlying mechanism being charge redistribution due to physical or chemical adsorption. These sensors function at room temperature, require relatively low input power, and can be highly sensitive. Their major drawback is poor gas specificity, due to the lack of spectroscopic identification of a particular species, so they may respond similarly to various analytes within a gas mixture.\\

Plasmon-enhanced infrared optical absorption based on 2D materials is regarded to be a promising spectroscopic technique for probing vibrational modes of large complex biopolymers (proteins, nucleic acids, synthetic polymers, or adsorbed molecular layers), as it enables high-fidelity mode sensing at room temperature \citep{avouris2017, yang2018, li2014, rodrigo2015, hu2016, oh2018a, farmer2016}. To be able to excite the 2D plasmon by direct optical excitation, nanoribbon arrays have been used \citep{ju2011, yan2013}. This technique, however, has not been proven as effective to identify the vibrational modes of gas molecules, where the main challenge is the small adsorption strength of the vibrational modes of individual molecules compared to biopolymers. However, the gas dielectric response is proportional to its concentration. This suggests that via trapping molecules close to 2D material, one might be able to enhance the interaction of gas molecules with the evanescent plasmon fields.\\

In a plasmon-based sensing setup, the analyte species may interact with the incident optical field \citep{jonas2008, yuan2017, zhang2010}, sensing material surface (via adsorption) \citep{kurihara2002, allsop2016}, and the applied bias field \citep{barik2014, oh2018, schafer2015, gascoyne2002}. These interactions have also been exploited in 2D-based platforms to control the positioning of nanoparticles \citep{engel2018, zhang2016, marini2015, vasic2015, barik2017}. In this work, we calculate the contributions of each of the three mechanisms (optical field, bias electrostatic field, and adsorption) in redistributing a homogeneous gas and thus facilitating gas sensing via plasmon excitation in a graphene nanoribbon (GNR) array sensor geometry. 

\section{Trapping Molecules}
The sensor geometry is illustrated in Fig. 1(a), where a simple metal-oxide-GNR is used to tune electrostatic doping in graphene. The gas is in direct contact with the GNR array. The incident light is impinging normally on the device with polarizations parallel and perpendicular to the GNR array. The reflected light is spectrally analyzed for gas signatures. The gas dielectric function is assumed to follow a Lorentzian form with frequency, $\omega$ within the spectral window of interest \citep{haug2009, rodrigo2016}:
\begin{equation}
\epsilon_{g} = 1 + \frac{\Delta \epsilon \,\, \Omega^{2}}{\Omega^2 - \omega^2 - i\gamma\omega},
\label{epsilon_gas}
\end{equation}

\begin{figure}
\centering
\includegraphics[width=\linewidth]{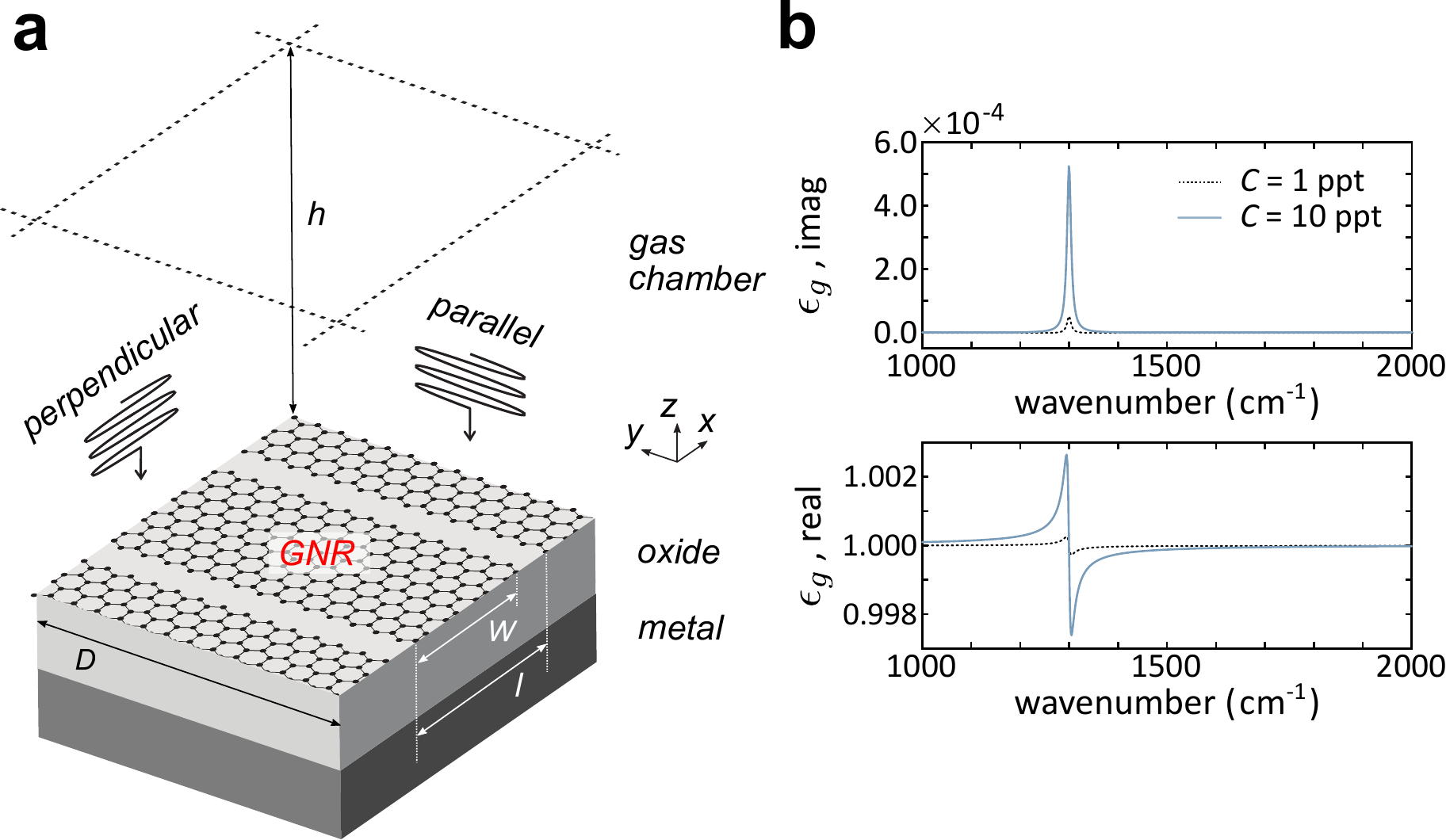}
\caption{(a) The schematic of the sensor setup. (b) The real and imaginary parts of gas permittivity vs wavenumber for 1 and 10 ppt gas concentrations under standard temperature-pressure condition.}
\label{schematic}
\end{figure}

\noindent where one IR-active vibrational mode is considered for the gas; the mode is characterized by its frequency $\Omega$, linewidth $\gamma$, and absorption weight $\Delta \epsilon$. We use $\Omega$ = 1300 cm$^{-1}$ and $\gamma$ = 10 cm$^{-1}$ as representative gas parameters. The absorption weight varies linearly with the gas concentration, i.e. $\Delta \epsilon = p_{0} \, \mathcal{C}$ \citep{liu2013, hu2019}, with the proportionality constant $p_{0}$ = 10 cm$^{3}$/mol \citep{hu2019}. The dielectric function in Eq. (1) is used for gas in both free and trapped states. The polarizability of the gas can then be obtained via \citep{griffiths2005}: 
\begin{equation}
\alpha_g = \frac{3\epsilon_{0}}{\mathcal{C}}\left(\frac{\epsilon_{g}-1}{\epsilon_{g}+2}\right),
\label{Lorentz-Lorenz}
\end{equation}

\noindent where $\epsilon_{0}$ is the vacuum permittivity. The gas dielectric function at 1 and 10 part per thousand (ppt) are shown in Fig. 1(b). Accordingly, the gas dielectric function can be tuned through its concentration, the peak magnitudes in both real and imaginary parts of $\epsilon_{g}$ vary linearly with concentration.\\

For gas molecules treated as classical particles in thermal equilibrium, the connection between local gas concentration, $\mathcal{C}(x,z)$ and trapping potential, $\mathcal{U}(x,z)$ can be obtained via statistical arguments (see Supplemental Material) \citep{kittel1998}:
\begin{equation}
\mathcal{C} = \frac{h\,l\,\mathcal{C}_{t} \left[1+\exp \left(-\beta \,\mathcal{U}\right)\Theta \left(-1-\beta \,\mathcal{U}\right)\right]}{h\,l + \int_{-\beta \,\mathcal{U} \, \geqslant 1} \exp\left(-\beta \,\mathcal{U}\right) dx dz}
\label{conc_gas}
\end{equation} 
where $\Theta(\cdot)$ is the step function, $h$ is the height of the gas chamber, and $l$ is the array period (see Fig. 1(a)). We consider $W =$ 50\,nm and $f = 0.7$ for the GNR width and width-to-period ratio, respectively. We note the setup has translational symmetry along $y$-coordinate, hence trapping potentials do not depend on $y$. The initial gas distribution is uniform with the concentration $\mathcal{C}_{t}$; unless denoted otherwise, 10\,ppt is taken for its value.\\

The last input is the trapping potential, to be written as a sum of two terms, one repulsive and the other attractive. The former is dominant when the molecule is in close vicinity to the surface and is modeled as an infinite potential for $z < z_{0}$ \citep{atkins1998}, with $z_{0}$ the equilibrium distance between the adsorbed molecule and graphene, and is comparable with molecule radius. The attractive term depends on the trapping mechanism involved and is to be developed for each mechanism separately. For adsorption, this is approximated via:
\begin{equation}
\mathcal{U}_{\textsf{ad}} = - D_{0} \exp\left[-\gamma_{0}(z-z_{0})\right], ~~~~ |x| < W/2
\label{Morse}
\end{equation}
which coincides in form with the attractive term in Morse potential commonly used to describe van der Waals interaction \citep{weinberger2016, pu2014}. $D_{0}$ denotes the binding energy of the molecule-2D material system, and $\gamma_{0}$ is related to the potential force constant close to $z_{0}$. Here, we assume: $\gamma_{0}$ = 10 nm$^{-1}$, $D_{0}$ = 0.3 eV, and $z_{0}$ = 0.3 nm \citep{liu2014a, zhang2009, lee2009}.\\

\begin{figure}[!b]
\centering
\includegraphics[width=\linewidth]{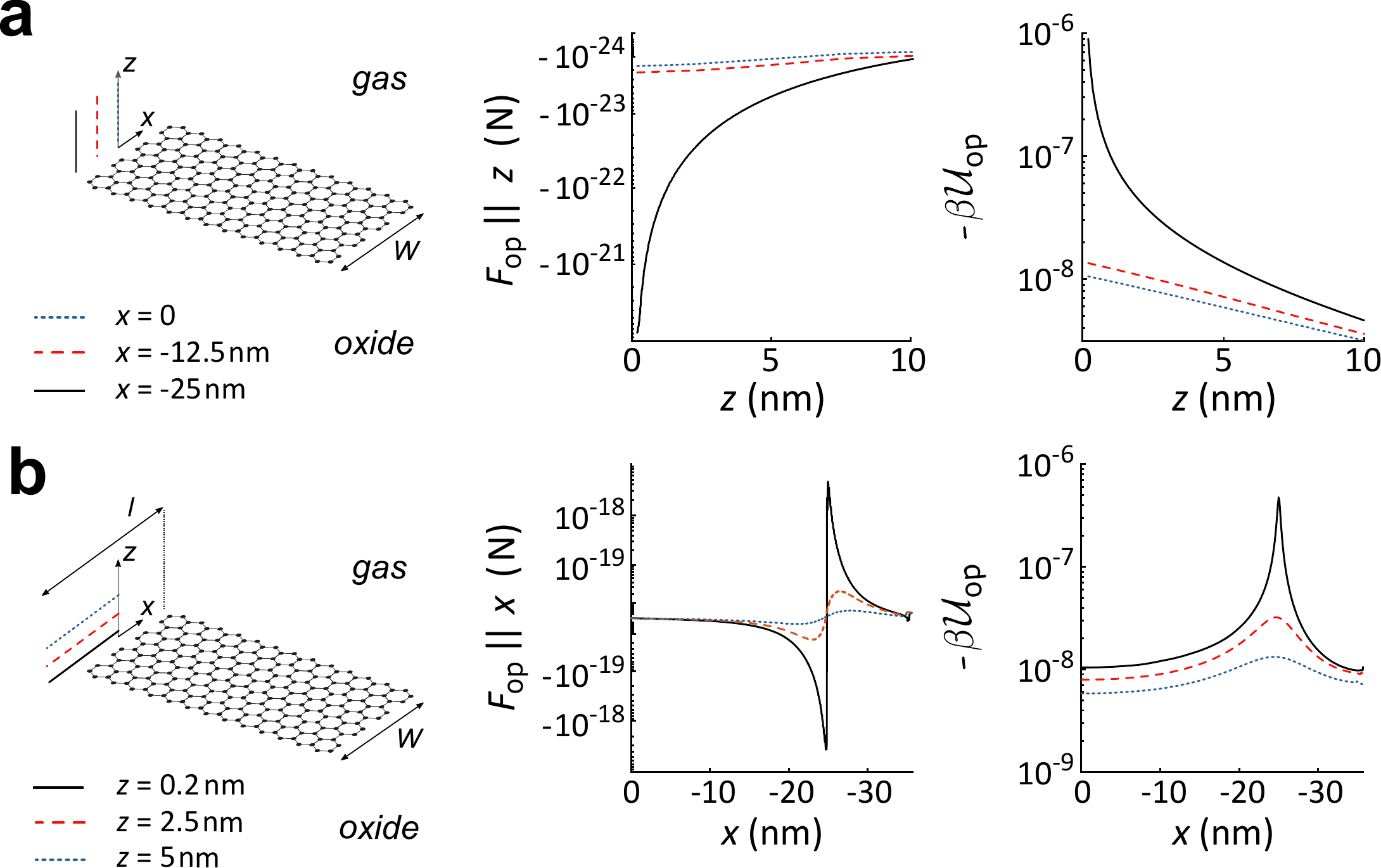}
\caption{(a) The $z$-component of optical force and optical potential energy along the cutlines parallel to $z$-axis, with $x$ = 0, -12.5, and -25\,nm. (b) The $x$-component of optical force and optical potential energy along the cutlines parallel to $x$-axis, placed 0.2, 2.5, and 5\,nm above graphene. The data are computed at 1295\,cm$^{-1}$ with 0.05\,kW/cm$^{2}$ as input power density.}
\label{opt_force}
\end{figure}

To determine the effect of the electrostatic gate bias, we solve for a self-consistent solution of Poisson's equation and the graphene net electron density \citep{barik2017}:   
\begin{equation}
\begin{split}
\vec{\nabla}. \left(\epsilon \vec{\nabla} \phi\right) & = e n(x)\delta(z)/\epsilon_{0},\\
n(x) & = \frac{1}{\pi} \left(\frac{\mu(x)}{\hbar v_{F}}\right)^2- n_{\textsf{D}}, ~~ |x| < W/2
\label{Poisson}
\end{split}
\end{equation}
where $e$ is the electron charge, $\phi(x,z)$ is the electrostatic potential, $v_{F} = 10^8$ cm\,s$^{-1}$ is the graphene Fermi velocity. $\epsilon$ is the spatially varying dielectric constant, which is assumed to be unity within the chamber and is equal to $\epsilon_{\textsf{ox}}\,$= 2.5 for the oxide region. $\mu(x)$ is the position-dependent chemical potential, which is given by: $\mu(x) = e\phi(x,z = 0)$. $n_{\textsf{D}}$ is the density of donor impurities in chemically doped graphene and is set to be 1.45 $\times$ 10$^{13}$ cm$^{-2}$. The problem is solved with COMSOL Multiphysics, AC/DC Module, where we applied the Dirichlet, periodic and Neumann boundary conditions for the bottom, sides and top edges of the simulation domain, respectively. The electrostatic force acting on a molecule is computed via: $\vec{F}_{\textsf{es}} = \frac{1}{2}\alpha_{0}\vec{\nabla} |\vec{E}|^2$ and the potential energy due to induced dipole is given by $\mathcal{U}_{\textsf{es}} = -\int\vec{F}_{\textsf{es}}.d\vec{r}$, where $\alpha_{0}$ = 10$^{-36}$ F.cm$^2$ is the static polarizability of a typical gas molecule \citep{duley1984, sircar2006}.\\

The optical force is calculated using, $\vec{F}_{\textsf{op}} = \frac{1}{4}\Re \{\alpha_{g}\} \vec{\nabla} |\vec{E}|^2$, where $\vec{E}$ is the sum of incident and reflected fields at a given point \citep{jonas2008, novotny2012}. The relation used here for optical force corresponds to its conservative component, known as gradient optical force \citep{novotny2012}. The scattering optical force, proportional to field phase gradient is negligible ($\sim\,$2 orders smaller in magnitude relative to gradient force for the parameters used in Fig. 2) and is disregarded throughout this work. Force calculations are performed using COMSOL Multiphysics, RF Module, and the corresponding potential energy is computed with, $\mathcal{U}_{\textsf{op}} = -\int\vec{F}_{\textsf{op}}.d\vec{r}$, implying that the optical potential energy is governed by the light intensity at a given point in chamber.\\

With the numerical recipes at our disposal, we proceed with the discussion of these trapping potentials and their relative abilities in trapping the free gas. We begin with the optical force and its potential energy, as summarized in Fig. 2. The ribbon edges constitute the hotspots for optical trapping. This is expected as the underlying plasmon field decays quickly away from the edges. For the parameters used in Fig. 2, the field amplitude at the edge is $\sim\,$20 times larger in magnitude and decays $\sim\,$145 times faster along the $z$-axis compared to those at the ribbon center (the 1/e amplitude drop in units of $\pi/W$ at the edge and center are 130 and 0.9, respectively), see Supplemental Material. Our simulations show the depth of the trapping potential is maximum when the plasmon resonance occurs near the gas mode resonance, i.e. when both the real part of gas polarizability and the field intensity at graphene surface are maximized. Moreover, the depth increases linearly with the intensity of the incident light. Despite this trend, however, for power intensities typically used in FTIR experiments (0.01-50 kW/cm$^2$), the maximum depth of the trapping potential is much smaller than the thermal energy at room temperature. Thus, for the geometry shown in Fig. 1(a), the field enhancement due to plasmon excitation does not help in trapping the molecules for sensing. We should point out, however, that the conclusion drawn here is for gas molecules atop graphene under plane wave illumination. The optical trapping force can be enhanced for molecules with larger $\Delta \epsilon$ under tightly focused light fields \citep{liu2011}, or for other 2D systems which exhibit in-plane hyperbolic response with larger plasmon-induced field confinement close to the material surface \citep{nemilentsau2016, van2018}.\\

\begin{figure}
\centering
\includegraphics[width=0.9\linewidth]{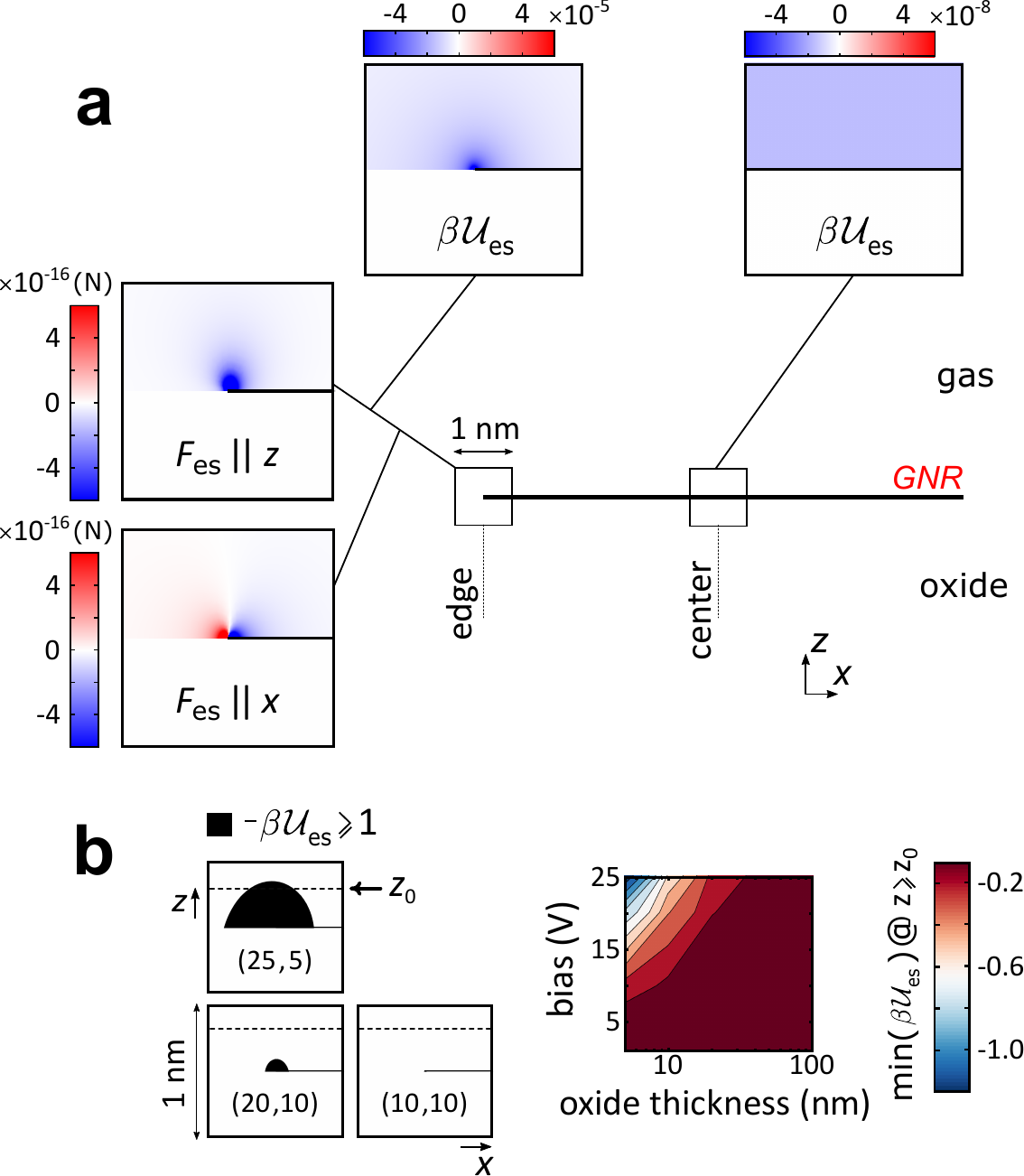}
\caption{(a) The spatial profile for electrostatic force components and potential energy in 1\,nm vicinity of ribbon edges. The potential energy at the ribbon center is also shown. The results are obtained for 10\,V bias and oxide thickness of 1.22\,$\mu$m. (b) Left: The region where the attractive electrostatic energy is smaller in magnitude relative to thermal energy. The data are shown for three pairs of (bias V, thickness nm). Right: The minimum electrostatic potential energy for $z\geq z_{0}$ as functions of bias and oxide thickness.}
\label{ads}
\end{figure}

We next look into electrostatic potential due to gate bias. From Fig. 3(a), the trapping hotspots again resides at the ribbon edges, and the potential energy decreases in magnitude rapidly towards the center, which is qualitatively similar to what we observed for the optical force. From Fig. 3(b), with increasing bias or reducing oxide thickness, the region where the electrostatic potential can surpass thermal energy becomes wider. For the assumed $z_{0}$ and with the repulsive potential included, 20\,V bias across 5-nm thick oxide are needed for the electrostatic potential to initiate gas trapping.\\

Lastly, we focus on surface adsorption and its effect on redistributing the gas molecules. In Fig. 4(a) the adsorption potential and the corresponding concentration of trapped molecules versus vertical distance from graphene surface are depicted. The inset shows the linear dependence of the surface density of adsorbed molecules, $n_{\textsf{ad}}$ as a function of the concentration of free molecules (proportional to gas pressure) in the chamber. The slope increases for a larger binding energy, which could occur in defective graphene, or via introducing chemical dopants \citep{zhang2009, dai2009, liu2010}. Our calculations show that for 5\,nm oxide thickness, 20\,V bias, and $z_{0}$ as small as 0.1\,nm, the maximum trapped concentration due to adsorption remains larger ($\sim$\,10$^{4}$) compared to that of the electrostatic trapping, which indicates that adsorption is the main mechanism that redistributes the free gas. 
 
\section{Plasmon-Based Sensing}
For gas modes to appear vividly in the sensor response one needs for the graphene plasmon resonance to match the gas characteristic mode while the quarter-wavelength condition is also satisfied. These criteria restrict both the oxide choice and oxide thickness ($\epsilon_{\textsf{ox}}\leq5$ and thickness of 0.5\,$\mu$m or more are needed), for typical background dopings ($n_{\textsf{D}}\lesssim 5\times10^{13}$\,cm$^{-2}$) in graphene \citep{lee2018, liu2011a}. We, therefore, focus in this section on how adsorption may facilitate gas sensing with graphene plasmons. We define the plasmon extinction as: 1-$R^{g}_{\textsf{per}}/R^{g}_{\textsf{par}}$, where $R^{g}_{\textsf{per}}$ and $R^{g}_{\textsf{par}}$ are the reflected powers when the chamber is filled with gas and the incident light is polarized along the $x$- and $y$-axes, respectively (for an alternate extinction definition, see Supplemental Material). The electromagnetic simulations are performed with COMSOL Multiphysics, RF module. The adsorption affects the sensor response through: (i) gas redistribution, which is incorporated via Eqs. \ref{epsilon_gas} and \ref{conc_gas} with the adsorption potential as input, we define accordingly an effective conductivity for the adsorbed layer: $\sigma_{\textsf{ad}} = -i\omega\epsilon_{0}h_{\textsf{ad}}\left(\epsilon_{g}(n_{\textsf{ad}}/h_{\textsf{ad}})-1 \right)$ \citep{hu2019}, where $h_{\textsf{ad}}$ denotes the height of the trapping region for which $-\beta \,\mathcal{U}_{\textsf{ad}}\, \geqslant 1$. (ii) doping through adsorption charge transfer. The latter is accounted for within the expression used for graphene conductivity \citep{gonccalves2016}: $\sigma_{\textsf{gr}} = ie^{2}v_{F}\sqrt{(n_{\textsf{D}} + n_{\textsf{ad}} Q)/\pi}/\left(\hbar\omega+i\eta\right)$, where $\eta$ = 50 meV is the broadening and $Q$ is the effective fractional charge transferred between graphene and one adsorbed molecule. For molecules which act as acceptors (donors) upon adsorption, $Q$ is negative (positive); in this work $Q$ = -0.05 is assumed \citep{kong2014}.\\

\begin{figure}
\centering
\includegraphics[width=\linewidth]{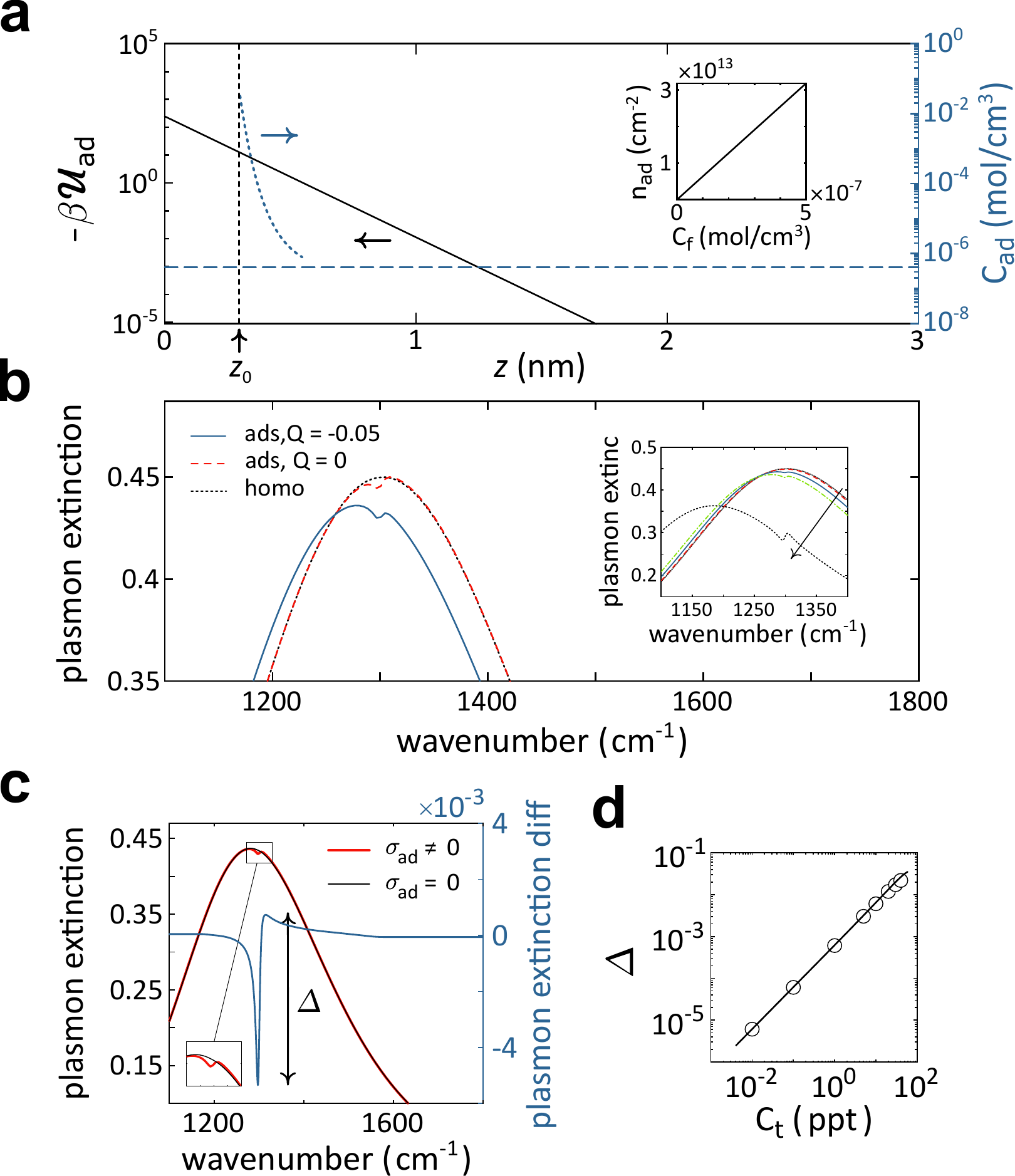}
\caption{(a) The adsorption potential and trapped gas concentration vs $z$. The horizontal dashed line denotes the $\mathcal{C}_{t}$. The inset shows the adsorbed surface density vs free gas concentration. (b) The plasmon extinction for homogeneous and adsorbed gas (with and without adsorption charge transfer). The inset
shows the plasmon extinction at various initial homogeneous gas concentration. Following the arrow $\mathcal{C}_{t}$ increases from 0.1, to 1, 5, 10, and 50 ppt. (c) The plasmon extinctions for $\sigma_{\textsf{ad}}$ included and when it set to zero. The corresponding plasmon extinction difference is also shown. (d) The variation amplitude in plasmon extinction difference vs gas homogeneous concentration. The solid line is obtained from the analytical model.}
\label{ads}
\end{figure}

To examine how adsorption modifies plasmon extinction, we begin with a comparison between the homogeneous and adsorbed gas distributions in Fig. 4(b). For graphene plasmon, adsorption in principle may shift the resonance either via doping or through modifying the dielectric function atop graphene. The latter, however, is relatively weak for all relevant adsorption densities. This is shown in Fig. 4(b) where the extinction peaks for homogeneous gas distribution and adsorbed gas with $Q$ = 0 (i.e. excluding adsorption charge transfer) are compared. Adsorption can affect the vibrational mode absorption strength in two ways, by tuning the gas concentration close to graphene (thereby modifying the mode weight) or by plasmon-gas mode resonance detuning. From the inset, it is clear that increasing the adsorbed density (via increasing $\mathcal{C}_t$ or $D_{0}$), results in more apparent gas-induced dips in the extinction. The inset data also imply that the mode weight is a more critical factor than detuning to control the dips amplitude in this setup.\\ 

We note that label-free plasmonic gas sensing schemes, commonly rely on direct chemisorption or physisorption of gas molecules atop the plasmonic surface \citep{shegai2012, hu2019}, with either wavelength-shift or intensity inspection adopted as the sensing method \citep{tittl2014, farmer2016}. The wavelength-shift, for example has been used for direct sensing of chemisorbed hydrogen on palladium plasmonic nanodisks \citep{wadell2014, tittl2011, tittl2012}. This metric, however, is not generally gas specific, which restricts wavelength-shift to situations in which sensing device is exposed to one gas only. In this work, the gas-graphene interaction involves physisorption. Moreover, we rely on intensity interrogation to achieve gas identification. In Fig. 4(c), the difference between plasmon extinctions when $\sigma_{\textsf{ad}}$ is included and when it is set to zero, are shown. The zero case denotes the extinction when the plasmon shift triggered by adsorption charge transfer is included while the contribution due to adsorbed gas dielectric response (which is the origin of the extinction dips) is left out. The peak-to-peak variation in the extinction difference, denoted with $\Delta$ is then recorded as a function of gas concentration in Fig. 4(d). From this, we conclude that for gas concentration of ppt level, plasmon extinction changes of the order 0.1\% can be observed, a level measurable by common far-field spectroscopic techniques \citep{hu2019}.\\

Finally, we note that a simple analytical model can be used to estimate the electromagnetic response of the sensor \citep{kotov2017, khaliji2017}. This is accomplished via introducing an equivalent conductivity for the total system of graphene ribbons and adsorbed molecules:
\begin{equation}
{\bf{\underline{\sigma_{t}}}}^{-1} = \frac{1}{f}\frac{1}{\sigma_{\textsf{gr}}+\sigma_{\textsf{ad}}}\,{\bf\underline{I}} + 
\begin{pmatrix}
-\frac{1}{i\omega \mathcal{C}_{c}} & 0\\
0 & 0
\end{pmatrix},
\label{cond_tot}
\end{equation}
where, $\bf{\underline{I}}$ is the unity matrix and $\mathcal{C}_{c} = \frac{l}{\pi}\epsilon_{0}\left[\epsilon_{\textsf{ox}}+\epsilon_{g}\left(\mathcal{C}_{t}-n_{\textsf{ad}}/h\right)\right]\log\left[\csc\left(\pi (1-f)/2\right)\right]$ is the coupling capacitance. Transfer matrix method outlined in ref. \citep{zhan2013}, is used to obtain the reflection spectra for the layered structure. From Fig. 4(d), the simple model recovers the magnitude change in gas-induced dips with the gas concentration.\\

\section{Concluding Remarks}
We examined the plasmonic sensing of gas vibrational modes using the graphene nanoribbon scheme. The sensitivity of gas detection depends on the trapping forces exerted on the gas molecules and by the confinement of the plasmonic field. These in turn also depend on the device structure. Here, through systematic modeling of the likely gas trapping mechanisms in experiments (optical forces, electrostatic forces and adsorption), we found that surface adsorption is the dominant mechanism in trapping free gas molecules atop graphene, which then enables plasmon enhanced sensing of the gas vibrational modes. Further increases in sensitivity are expected by utilizing a perfect absorption scheme \citep{thongrattanasiri2012, kim2018, guo2018} or the recently described approach \citep{lee2019, iranzo2018, alonso2017, chen2017} of placing an underlayer metal in close proximity to graphene to exploit extreme plasmon confinement due to the excitation of acoustic plasmons.

\section*{Acknowledgments}
This research was supported by grants from the U.S. National Science Foundation (ECCS 1809723 to T.L. and S.-H.O. and ECCS 1809240 to S.-H.O.). S.-H.O. also acknowledges support from Seagate through the MINT consortium at the University of Minnesota.


\end{document}